\documentclass[prl,showpacs
,twocolumn]{revtex4}
\newcommand{\be}{\begin{equation}} 
\newcommand{\ee}{\end{equation}} 
\newcommand{\bea}{\begin{eqnarray}} 
\newcommand{\eea}{\end{eqnarray}}

\usepackage{graphicx}
\usepackage{bm}

\begin{document}
\title{Statistics of Intense Turbulent Vorticity Events}
\author{L. Moriconi}
\affiliation{Instituto de F\'\i sica, Universidade Federal do Rio de Janeiro \\C.P. 68528, Rio de Janeiro, RJ --- 21945-970, Brasil}
\begin{abstract}
We investigate statistical properties of vorticity fluctuations in fully developed turbulence, which are known to exhibit a strong intermittent behavior. Taking as the starting point the Navier-Stokes equations with a random force term correlated at large scales, we obtain in the high Reynolds number regime a closed analytical expression for the probability distribution function of an arbitrary component of the vorticity field. The central idea underlying the
analysis consists in the phase-space restriction to a particular sector where the rate of strain and the rotation tensors can be locally regarded as slow
and fast degrees of freedom, respectively. This prescription is implemented along the Martin-Siggia-Rose functional framework, whereby instantons and
perturbations around them are taken into account within a steepest-descent approach.
\end{abstract}
\maketitle 

The intermittent nature of turbulent flows, long ago discovered by Batchelor and Townsend \cite{batch}, has been an issue of main concern in fluid dynamic research. Despite the great investigation efforts taken so far, boosted in the last two decades by considerable experimental and numerical progress, there is not yet a clear theoretical understanding of intermittency (see Ref. \cite{sreenv1} for an overview). In fact, the sudden and intense vorticity bursts observed at high Reynolds numbers can be addressed in general along two distinct -- but not necessarily incompatible -- viewpoints. Strong fluctuations may arise as a consequence of local vorticity amplification by stretching, induced on its turn by strain defined at larger scales. Alternatively, high peaks of enstrophy and dissipation are also detected whenever a vortex coherent structure is advected across the measurement position. Such an interpretation ambiguity was emphasized in the recent experimental work by Zeff et al. \cite{zeff}, where some support was given for the dominance of the former, local mechanism of vorticity bursting.

Intermittency is frequently characterized in homogenous isotropic turbulence by the statistical properties of the velocity derivatives 
$s_{ij} \equiv \partial_i v_j$, which, as expected on general grounds, have their second and higher moments diverging as some power of the inverse viscosity \cite{frisch}. A relevant problem is to find the profile of the probability distribution function tails (pdf tails) of $s_{ij}$. Exact answers have been found in the simpler realm of the Burgers model, where the presence of shock waves leads to a bifractal form of intermittency \cite{aurell}. It turns out that in general the velocity derivative pdf tail converges in a pointwise sense to $C|s|^{-7/2}$ \cite{we,bec}, for $s<0$, in the vanishing viscosity limit $\nu \rightarrow 0$ (for $s>0$ the pdf decay is faster than gaussian). In contrast, much less is known in the case of three-dimensional incompressible turbulence. Actually, just a few attempts to describe the viscous asymptotic expressions for the pdfs of the velocity derivatives are available so far, devised on the grounds of phenomenological multifractal ideas \cite{frisch2,benzi,arimitsu}, which stress the dissipative role played by small scale vortex structures.

The main result of this letter, to be obtained with the help of functional techniques, is an analytical expression for the tails of the vorticity pdf $\rho(\omega)$ in the high Reynolds number limit, where $\omega$ is an arbitrary cartesian component of the vorticity vector $\omega_i = \epsilon_{ijk} \partial_j v_k$. Our starting point is the effective modeling of turbulence, as described by the stochastic Navier-Stokes equations,
\be
\partial_t v_\alpha + v_\beta \partial_\beta v_\alpha = 
-\partial_\alpha P + \nu \partial^2 v_\alpha + f_\alpha \ , \  \partial_\alpha v_\alpha = 0 \ . \ \label{st-ns} 
\ee
The gaussian random force $f_\alpha (\vec x, t)$ has a vanishing mean value and the two-point correlator
$D_{\alpha \beta}(\vec x-\vec x') \delta (t - t')$, with
\be
D_{\alpha \beta}(\vec x-\vec x') =  \delta_{\alpha \beta}
D_0 \exp (-|\vec x - \vec x'|^2 / L^2 ) \ . \
\ee
Assuming the Kolmogorov cascade picture \cite{frisch}, energy is injected into the fluid at rate $D_0$, within large length scales of order $L$, and dissipated at the microscopic scale $\eta \sim \nu^{3/4} \rightarrow 0$, where viscous effects come into play. 

Let $\Gamma=\oint_c d x_\alpha v_\alpha$ be the circulation evaluated at time $t=0$, for a circular contour $c$ of radius $R \rightarrow 0$, counterclockwise oriented and centered at the origin in the $xy$ plane. The vorticity pdf $\rho( \omega)$ may be written in terms of the characteristic functional $Z( \lambda) \equiv \langle \exp(i \lambda \Gamma ) \rangle$ as
\be
\rho(\omega)=\lim_{R \rightarrow 0} \frac{R^2}{2 \pi} \int_{- \infty}^{\infty} d \lambda \exp(-i \lambda \omega R^2) Z(\lambda) \ . \ \label{vpdf-def}
\ee
The radius $R$ enters in the formalism as a regularizing parameter to handle some singular expressions which appear in the course of computations ($R \rightarrow 0$ is taken at fixed $\nu$). The Martin-Siggia-Rose functional approach \cite{msr} may be now recalled to put $Z(\lambda)$ into a path integration form. Up to a normalization factor \cite{comment}, we have
\be 
Z(\lambda)= \int D \hat v Dv DP DQ \exp  ( i S ) \ , \ \label{zmsr} 
\ee 
where the Martin-Siggia-Rose action is given by 
\be
S =\int d^3 \vec x \int_{-T}^{0} dt {\cal{L}} + \lambda \Gamma \ , \ \label{smsr}
\ee
related to the non-local lagrangian density
\bea 
{\cal{L}} &=&
\hat v_\alpha (\partial_t v_\alpha + v_\beta  \partial_\beta  v_\alpha -\nu \partial^2 v_\alpha 
+ \partial_\alpha P) + Q \partial_\alpha v_\alpha  \nonumber \\ 
&+& \frac{i}{2} \hat v_\alpha \int d^3 \vec x' D_{\alpha \beta}(\vec x - \vec x') \hat v_\beta (\vec x', t)  \ . \ \label{lmsr} 
\eea
Of course, the limit $T \rightarrow \infty$ is intended in (\ref{smsr}). However, we will resort in practice to a temporal domain of size $T =\kappa L^2/ \nu$, having the order of the viscous decay time of smooth large scale fluctuations (introduced in the forthcoming instanton computations), where $\kappa$ is some Reynolds number dependent dimensionless parameter. Free boundary conditions are allowed at $t= -T,0$.

As an attempt to single out the phase-space sector where strong fluctuations of $\Gamma$ develop, we perform the analytical mapping 
$\lambda \rightarrow -i \lambda$ in (\ref{smsr}), taking afterwards the large $\lambda$ limit. The saddle-point technique may then be
employed to set non-perturbative field configurations (instantons) which would solve the variational equation $\delta S =0$, and hopefully yield the dominant asymptotic behavior of $Z(\lambda)$, directly related to the form of the vorticity pdf tails. These computational steps, which can be carried out for the analysis of pdf tails of general observables, constitute the essence of the instanton method \cite{falko}, successfully applied to problems like the random advection of a passive scalar and Burgers turbulence \cite{falko,gura,chertkov,balko}. Unfortunately, the implementation of the instanton approach in the three-dimensional incompressible situation is plagued with non-perturbative difficulties, as previously noted in Ref. \cite{falko}. It is somewhat simple to understand why the saddle-point action that would be obtained along the above lines cannot lead to the correct vorticity pdf tails (or to the pdf tails of any linear combination of velocity derivatives). Since the saddle-point equations are invariant under the transformations
\bea
&&\nu' = h^{1/2} \nu \ , \  
t' = h^{-1/2} t \ , \ 
x'=x \ , \
v_\alpha' = h^{1/2} v_\alpha \ , \ \nonumber \\
&&\hat v_\alpha' = h \hat  v_\alpha \ , \
P' = h P \ , \
Q' = h^{3/2} Q \ , \ 
\lambda' = h \lambda \ , \  \label{sc-transf} 
\eea
where $h$ is an arbitrary scaling parameter, it follows that the saddle-point action has necessarily the general form $S = \lambda^{3/2} f(\lambda^{-1/2} \nu)$, yielding, in the high Reynolds number limit, $\ln [ \rho(\omega) ] \sim - \omega^3 $ at the rigth tail (that is, $\omega >0$), which is a very unlikely expression under the light of numerical and experimental findings \cite{vincent,sreenv2,tabeling,farge}.

One might conjecture that the functional approach would work if fluctuations around the instanton were taken into account, possibly to all orders in the perturbation series, to overcome the $\lambda^{3/2}$ dependence in the saddle-point action. However, this is an extremely hard task to accomplish. A more computationally oriented point of view is to hypothesize that a family of instanton solutions could be defined from the start, so that the overall contribution of the associated saddle-point actions would lead ultimately to physically meaningful results. Indeed, we have found that the latter option is
the most promising one. 

The hint on how to set up a family of instanton configurations comes from the hypothesis that the small scale, intense bursts assumed to dominate the inviscid vorticity pdf tail are driven by the rate of strain provided by larger eddies, which according to standard ideas, fluctuate on slower time scales \cite{teneekes}. In fact, several turbulence models depict vorticity enhancement along the energy cascade as a local process which takes place in a static background defined by an external straining flow \cite{pullin}. Among the manifold scenarios proposed in the literature, it is worth mentioning Lundgren's modeling of stretched vortex tubes with internal spiral structure \cite{lundgren}, intimately connected with the observations reported in a remarkable recent experiment by Cuypers et al. \cite{cuypers}. 

The above phenomenological considerations motivate us to revisit the functional strategy initially founded on Eqs. (\ref{vpdf-def}-\ref{lmsr}). The point is to separate, in the integration measure, the slow degrees of freedom (the rate of the strain tensor) from the fast ones (the rotation tensor). To achieve this goal, we insert the identity $1=\int D \sigma^s \delta [ \sigma^s_{\alpha \beta} -( \partial_\alpha v_\beta + \partial_\beta v_\alpha)/2 ]$
in the integrand of (\ref{zmsr}). Defining
\be
\tilde Z([ \sigma^s ], \lambda) =  \int D \hat v Dv DP DQ D \tilde Q \exp(i \tilde S) \ , \ \label{ztmsr} 
\ee
a strain-dependent characteristic functional, associated to the modified Martin-Siggia-Rose action
\be 
\tilde S = S + \frac{1}{2}\int d^3 \vec x dt \tilde Q_{\alpha \beta}
\left (  \partial_\alpha  v_\beta  
+ \partial_\beta v_\alpha  - 2 \sigma^s_{\alpha \beta} \right) \ , \ \label{stmsr} 
\ee 
we write, after exchanging the order of integrations, the alternative (and exact) expression for the vorticity pdf in the conditional form,
\be
\rho(\omega)=\lim_{R \rightarrow 0} \frac{R^2}{2 \pi} \int D \sigma^s \int_{- \infty}^{\infty} d \lambda \exp(-i \lambda \omega R^2) \tilde 
Z([ \sigma^s ], \lambda) \ . \ \label{vpdf-defb}
\ee
It is clear, thus, that instanton configurations may be naturally labelled (in a functional sense) by the rate of the strain tensor $\sigma^s_{\alpha \beta}$, if they are now found as solutions of the saddle-point equations derived from $\delta \tilde S = 0$, viz.,
\bea
&&\partial_\alpha v_\alpha = \partial_\alpha \hat v_\alpha =\partial_\alpha v_\beta + \partial_\beta v_\alpha - 2 \sigma^s_{\alpha \beta}= 0  \ , \ \nonumber  \\
&&\Delta_+ \hat v_\alpha - \hat v_\beta \partial_\alpha 
v_\beta + \partial_\beta  ( v_\beta \hat v_\alpha +Q^\star_{\alpha \beta } ) + i \lambda  
{{\delta \Gamma} \over {\delta v_\alpha}} 
= 0 \ , \ \nonumber \\
&&\Delta_- v_\alpha + v_\beta \partial_\beta 
v_\alpha + \partial_\alpha P + i D_{\alpha \beta} \otimes \hat v_\beta =0 \ . \ \label{sp-eqns} 
\eea
Above, we have $\Delta_{\pm} \equiv \partial_t \pm \nu \partial^2$, $ Q^\star_{\alpha \beta } \equiv
 \tilde Q_{\alpha \beta }+ Q \delta_{\alpha \beta}$, and 
\be
\frac{\delta \Gamma }{\delta v_\alpha} =\epsilon_{3 \beta \alpha} {x_\beta \over r_\perp} \delta (r_\perp-R) \delta(z) \delta(t) \ , \ \label{funct-der} 
\ee 
where $r_\perp =\sqrt{x^2+y^2}$. As can be easily checked, the saddle-point Eqs. (\ref{sp-eqns}) are invariant under (\ref{sc-transf}), replacing $Q$ by $Q^\star_{\alpha \beta}$ and adding $(\sigma^s_{\alpha \beta})'= h^{1/2} \sigma^s_{\alpha \beta}$ to the set of transformations. At high Reynolds numbers,
the modified saddle-point action will have then the functional form $\tilde S = \lambda^{3/2}f[\lambda^{-1/2} \sigma^s]$.

It would not be an easy matter to solve exactly Eqs. (\ref{sp-eqns}). However, as a general and valuable feature of the instanton method, it is not necessary to know the detailed global form of solutions of the saddle-point equations \cite{falko,chertkov,gura,balko}. In our particular problem, the small radius $R$ of the integration contour for $\Gamma$, which is in the viscous range, indicates that a linearization of $v_\alpha (\vec x,t)$ around $\vec x = 0$ would suffice. Also, as a first order approximation, the rate of the strain will be considered time-independent, as suggested from its role as a slowly fluctuacting variable in the vorticity intensification process. 

While we are not worried with the global spatial behavior of the velocity field, the instantons are required to have finite action, and to satisfy the boundary condition $\hat v_\alpha (\vec x, 0^+) =0$, assuring that solutions for $\hat v_\alpha (\vec x,t)$ will not blow up for $t \rightarrow \infty$,  
once the parabolic operator $\Delta_{+}$ is non-negatively eigenvalued. We may replace, equivalently, the $\hat v_\alpha (\vec x, 0^+) =0$ condition by
\be
\hat v_\alpha (\vec x, 0^-) = i \lambda \epsilon_{3 \beta \alpha} {x_\beta \over r_\perp} \delta (r_\perp-R) \delta(z) \ , \ \label{bd-cond}
\ee
as it follows from the evolution equation for $\hat v_\alpha (\vec x,t)$ and the former boundary condition itself. Taking now $\partial_\beta Q^\star_{\alpha \beta} = \hat v_\beta (\partial_\alpha v_\beta - \partial_\beta v_\alpha )$ in (\ref{sp-eqns}), we get, for $t <0$,
\be
\Delta_+ \hat v_\alpha - \hat v_\beta \partial_\beta 
v_\alpha + v_\beta \partial_\beta \hat v_\alpha = 0 \ , \ \label{hatv-eqn}
\ee
which is a Helmholtz-like equation. It is in fact always possible to obtain a symmetric tensor solution for $Q^\star_{\alpha \beta}$, closely related to the result that the support of $\hat v_\alpha (\vec x,0^-)$, a singular ring, is backwardly advected in time by the reversed velocity field, as implied by Eq. (\ref{hatv-eqn}). It is not difficult to prove that the backward advection of the singular ring will generate a contour with bounded length (and hence a non-vanishing saddle-point contribution) for all $t<0$ if and only if $\sigma^s_{13} = \sigma^s_{23} =0$, and the two eigenvalues of $\sigma^s_{\alpha \beta}$ in the $xy$ plane are positive. Therefore, it is worth noting that the functional integration in (\ref{vpdf-defb}) turns to be effectively replaced by ordinary integrations over the linearly independent matrix components $\sigma^s_{11}$, $\sigma^s_{12}$ and $\sigma^s_{33}$ of the rate of the strain. Observe that $\sigma^s_{22}$ is not in the set of integration variables, since $\sigma^s_{22}+\sigma^s_{11}+\sigma^s_{33} =0$, according to the incompressibility condition. 

We focus the analysis of the saddle-point equations on the class of axisymmetric instantons, having in mind the axial symmetry of 
$\tilde Z([ \sigma^s ],\lambda)$. The general form of $\tilde Z([\sigma^s],\lambda)$ will be found through a symmetrization procedure, taking into account that $\tilde Z([\sigma^s],\lambda=0)$ is invariant under transformations of the full rotation group, and any eventual $\lambda$-dependent correction factor will be invariant only under rotations around the $z$ axis. In concrete terms, we are just taking advantage of the fact that Eqs. (\ref{sp-eqns}) become fully axisymmetric and consequently more manageable if we write the rate of the strain tensor -- the static background in the saddle-point equations -- as $\sigma^s_{\alpha \beta} = a(\delta^\perp_{\alpha \beta} -2 \delta_{\alpha 3} \delta_{\beta 3})$, where $a$ is an arbitrary positive parameter, and $\delta^\perp_{\alpha \beta} \equiv \epsilon_{3 \alpha \gamma} \epsilon_{3 \beta \gamma}$. We get, then, the saddle-point configuration
\bea
&&\tilde Q_{\alpha \beta} (\vec x ,t) = -i \lambda \omega(t) \delta^\perp_{\alpha \beta} \Theta(R \exp(at)-r_\perp) \delta(z) \ , \ \nonumber \\
&&Q (\vec x ,t)= 0 \ , \ P(\vec x ,t)=(\omega^2(t)/4 -a^2) r^2_\perp /2 - 2 a^2 z^2 \ , \ \nonumber \\
&&v_\alpha (\vec x ,t)= \sigma^s_{\alpha \beta} x_\beta - \omega(t) \epsilon_{3 \alpha \beta} x_\beta /2\ , \ \nonumber \\
&&\hat v_\alpha (\vec x ,t)= i \lambda \epsilon_{3 \beta \alpha} \frac{x_\beta}{r_\perp} \delta(r_\perp - R \exp(at)) \delta(z) \ , \ \label{sp-conf}
\eea 
where $\Theta( \cdot )$ is the Heaviside function, and
\be
\omega(t) = \frac{\pi D_0 \lambda R^2}{a L^2} \exp(-2 a |t|) 
\ee
is the instanton vorticity. The imposition of $v_\alpha(0,t)=0$ in (\ref{sp-conf}) is guaranteed from the invariance of $\tilde S$ under generalized galilean transformations \cite{comment2}, while the pressure $P(\vec x ,t)$ is fixed up to some unimportant additive time-dependent function, associated to the large scale boundary conditions. In order to consider the effects of smooth fluctuations around the instanton solutions, we take the fields given in (\ref{sp-conf}) and perform the substitutions $v_\alpha \rightarrow v_\alpha + \delta v_\alpha$, $\hat v_\alpha \rightarrow \hat v_\alpha + \delta \hat v_\alpha$, etc. in the path integration (\ref{ztmsr}), keeping in $\tilde S$ terms defined up to the second order in the perturbations. The integrations over $\delta \tilde Q$ and $\delta Q$ enforce the constraints $\partial_\alpha \delta v_\beta + \partial_\beta \delta v_\alpha = 0$ and $\partial_\alpha \delta v_\alpha = 0$, respectively. Therefore, still working in the linear approximation for the velocity field, we will have $\delta v_\alpha (\vec x,t) =\Phi_{\alpha \beta} x_\beta$, with $\Phi_{\alpha \beta} \equiv \epsilon_{\alpha \beta \gamma} \phi_\gamma$. Defining, furthermore, $\hat v_{\alpha \beta}(t) = \int d^3 \vec x x_\alpha \delta \hat v_\beta$ (an antisymmetric tensor, as it follows from the integration over $\delta P$), we obtain the phase-space reduced expression
\be
\tilde Z([\sigma^s],\lambda) = \exp(-\frac{\pi^2 D_0 R^4 \lambda^2}{2 L^2 a}) \int D \hat v_{\alpha \beta} D \Phi_{\alpha \beta} \exp(-M) \ , \
\ee
which depends on the effective action
\be
M=\int_{-T}^0 dt [ \frac{D_0}{L^2} \hat v_{\alpha \beta}^2 - i \hat v_{\alpha \beta} ( \dot \Phi_{\alpha \beta}
+\frac{1}{2} \sigma^s_{\gamma \delta} \epsilon_{\gamma \alpha \beta} \epsilon_{\delta \rho \eta} \Phi_{\rho \eta}) ] \ . \
\ee
The subsequent integration over $\hat v_{\alpha \beta}$ gives
\be
\tilde Z([\sigma^s],\lambda) = \exp(-\frac{\pi^2 D_0 R^4 \lambda^2}{2 L^2 a}) \int D \phi_\alpha \exp(-H) \ , \ \label{ztmsr-H}
\ee
where we have now
\be
H = \frac{L^2}{2 D_0} \int_{-T}^0 dt ( \dot \phi_\alpha - \sigma^s_{\alpha \beta} \phi_\beta)^2 + \epsilon [ \phi_\alpha^2(0)+\phi_\alpha^2(-T)] \ . \
\label{H}
\ee
The functional determinant factor in (\ref{ztmsr-H}) can be exactly computed, from its relation to the problem of a set of three one-dimensional euclidean harmonic oscillators \cite{parisi}. The term proportional to $\epsilon \rightarrow 0^+$ in $H$ provides a zero-mode regularization of the functional integration. We find, thus, taking together the saddle-point action and the fluctuations around the instantons,
\be
\tilde Z([\sigma^s],\lambda) =\frac{\exp( \pi^2 D_0 R^4 \lambda^2 / L^2 \sigma^s_{33})}
{\sqrt{{\hbox{det}} [ \cosh(\sigma^s T)]}} \ . \ \label{ztilde}
\ee
The rate of the strain tensor which appears in (\ref{ztilde}) can be diagonalized through a simple rotation by an angle $\theta$ around the $z$ axis. Let $\sigma^s_{11} = p-q$, $\sigma^s_{22}=p+q$ and $\sigma^s_{33} = -2p$ be the matrix elements of $\sigma^s$ in its diagonal form, where $p$ and $q$ are real parameters. The jacobian of the mapping $ (p,q,\theta) \rightarrow (\sigma^s_{11}, \sigma^s_{12}, \sigma^s_{33})$ is $4|q|$, and we write, on the basis of 
(\ref{vpdf-defb}),
\bea
\rho(\omega)=\lim_{R \rightarrow 0} \frac{R^2}{2 \pi} \int_0^\infty dp \int_0^{p} q dq \int_{-\frac{\pi}{2}}^{\frac{\pi}{2}} d \theta \nonumber \\
\times \int_{- \infty}^{\infty} d \lambda \exp(-i \lambda \omega R^2) \tilde 
Z([ \sigma^s ], \lambda) \ . \ \label{vpdf-red}
\eea
In the large $T$ limit the hyperbolic cosine in (\ref{ztilde}) is well approximated by a simple exponential and the vorticity pdf may be written in a closed analytical form. We get the t-distribution
\be
\rho(\omega) = \frac{15 (g \langle \omega^2 \rangle )^3}{16 (\omega^2 + g \langle \omega^2 \rangle)^{\frac{7}{2}}} \ , \ \label{vpdf}
\ee
where $g \propto \kappa$ and $\langle \omega^2 \rangle$ is the actual expectation value of the squared vorticity, to be taken from experiments or numerical simulations (we have $\int d\omega \omega^2 \rho(\omega) = g \langle \omega^2 \rangle /4$). We expect Eq. (\ref{vpdf}) to apply only at the pdf tails and their leading order corrections, where $|\omega| \gg \sqrt{\langle \omega^2 \rangle}$.





An excellent agreement holding around three decades of the vorticity pdf is found between the tails predicted by (\ref{vpdf}), with $g \simeq 5.5$
and $\sqrt{\langle \omega^2 \rangle} \simeq 10$, and the results of numerical simulations at moderately high Reynolds number \cite{farge}, as shown in Fig.1. The relevant vorticity range is localized considerably beyond the usual gaussian core. It is an interesting problem, left for further investigation, to determine the Reynolds number dependence, if any, of the fitting parameter $g$.

To summarize, we studied the form of vorticity pdf tails at high Reynolds numbers by means of the instanton technique. We took into account, as an essential phenomenological motivation, the idea that the intermittent vorticity bursts observed in turbulent flows arise from the coupling of vorticity and relatively slow fluctuations of the rate of the strain tensor. A supporting comparison was established between the vorticity pdf tails given by (\ref{vpdf}) and previous numerical results.

I thank M. Farge for providing numerical vorticity pdf data files. This work has been partially supported by FAPERJ.

FIG. 1:
The vorticity pdf tails taken from (\ref{vpdf}) (solid lines) are compared to the results of numerical simulations
by Farge et al. \cite{farge} at the Taylor-scale Reynolds number $R_\lambda = 150$.

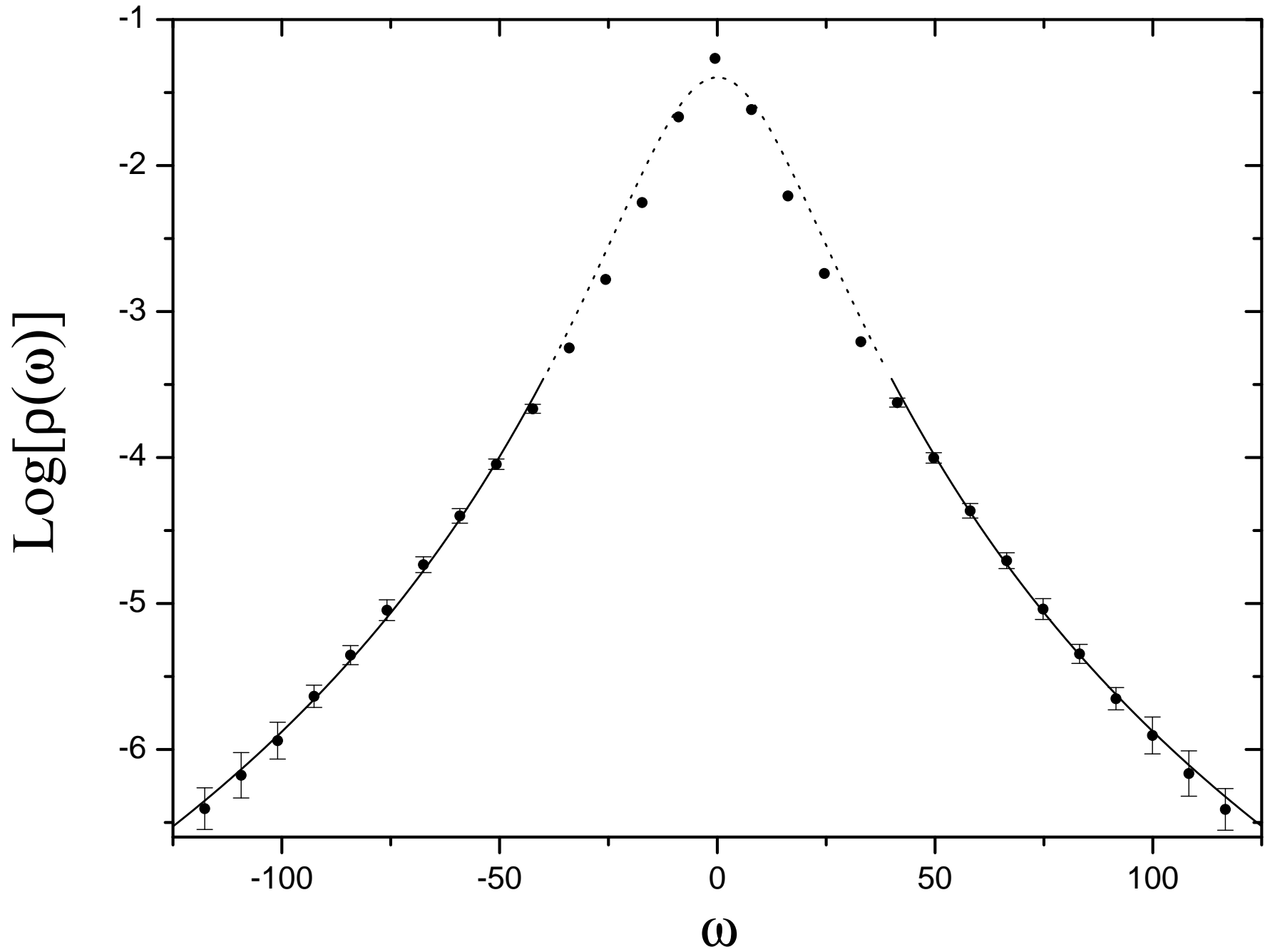
\end{document}